\documentclass[12pt, onecolumn]{article}
\usepackage{graphicx}
\usepackage{hyperref}

\title{The Art of Memory and the Growth of the Scientific Method\footnote{The title is adapted from Chapter 10 of Frances Yates' book \emph{The Art of Memory}.}}
\author{Gopal P. Sarma\footnote{Email:gsarma@alumni.stanford.edu}}
\date{}

\begin{document}
\maketitle
\begin{abstract}
I argue that European schools of thought on memory and memorization were critical in enabling the growth of the scientific method.  After giving a historical overview of the development of the memory arts from ancient Greece through 17th century Europe, I describe how the Baconian viewpoint on the scientific method was fundamentally part of a culture and a broader dialogue that conceived of memorization as a foundational methodology for structuring knowledge and for developing symbolic means for representing scientific concepts.  The principal figures of this intense and rapidly evolving intellectual milieu included some of the leading thinkers traditionally associated with the scientific revolution; among others, Francis Bacon, Renes Descartes, and Gottfried Leibniz.  I close by examining the acceleration of mathematical thought in light of the art of memory and its role in 17th century philosophy, and in particular, Leibniz' project to develop a universal calculus.
\end{abstract}

``What is the scientific method?''  It is a question that rarely makes an appearance in the scientific world.  And perhaps for good reason.  Most scientific disciplines have advanced so far, that to the extent that there is any ``method'' for students to learn, it is largely implicit knowledge that is absorbed by actively problem solving and participating in the frontiers of research.  \\

According to the Oxford-English Dictionary, the scientific method is ``a method of procedure that has characterized natural science since the 17th century, consisting in systematic observation, measurement, and experiment, and the formulation, testing, and modification of hypotheses.''\footnote{Scientific method [Def. 1]. (n.d.). Oxford English Dictionary Online. In Oxford English Dictionary. Retrieved May 16, 2013, from http://www.oed.com.}  There is no doubt that good science is characterized by the qualities encapsulated in the above definition, and as I stated above, that these qualities are learnt implicitly by participating in the scientific process under the guidance of more experienced researchers.  And yet, for an idea that is so simply stated, there seems to be a profound sense of mystery surrounding the historical events that allowed for the scientific method to take root at an institutional level.  What was it that took place in Europe during the 17th century that gave rise to a cultural transformation so significant that it changed the face of not only one area of systematic inquiry, but the entirety of the human pursuit of knowledge?  And how did a philosophical transformation of massive proportions happen during this period and why did it not happen before?  The world had certainly seen systematic scientific reasoning and hypothesis-driven investigation prior to the 17th century.  Empires had existed for thousands of years, and the very operational foundation of these multi-national entities would have required advanced knowledge to maintain such vast infrastructure.  We can point to numerous examples of sophisticated understanding in cosmology, medicine, or mathematics that are clear evidence of some form of principled reasoning that existed prior to the 17th century.  \\

And yet, it does seem that something profound happened during this time period that deserves explanation.  As modern intellectuals, we are surrounded by numerous forms of print and electronic media that allow us to participate in a philosophically reflectively culture that might shape our world views gradually.  We can read and re-read articles in newspapers, scientific journals, books, and magazines, we can watch and re-watch lectures on the Internet, and followup these periods of intense investigation with rapid electronic discussion with colleagues or anonymous commentators on blogs and question answering sites.  And to nurture ``pre-scientific'' or ``meta-scientific'' knowledge, elite academic institutions have created exclusive departments for philosophy and history of science.  \\

Almost none of these outlets and intellectual infrastructure existed in the 17th century, and certainly not in the period leading up to it.  What then would have enabled a highly philosophical stance toward the pursuit of knowledge, a ``meta-idea,'' to take root?  Even in the modern world, philosophically minded young people are chastened by their more worldly and grounded peers to try to focus on concrete problems.  Theoretical physicists lament the towers of abstraction built by mathematicians and engineers often hold a similarly derisive attitude towards their colleagues in physics.  In a world without all of these distinctions, what would have been the motivating forces for adopting a philosophical and abstract perspective with regards to the rational pursuit of knowledge, and not simply in the hands of a few brilliant visionaries but rather at an institutional and cultural level?  Why did these perspectives gain momentum and why had they not before?  The profoundly significant differences between our own world-- even the recent world prior to the Internet-- and European intellectual climate several centuries ago suggests that perhaps there were important factors that have been forgotten.  \\

Why should we care to answer these questions?  Other than the intrinsic historical value, there are perhaps more pressing contemporary concerns.  We live in an age in which the scope of science is larger than it has ever been, in which more scientists are being trained than ever before, and in which the tectonic plates of the scientific establishment are shifting, with such headlining trends such as the massive, multi-institution, multi-national ``big science'' projects, or the petabytes upon petabytes of data being produced which demand a global computational infrastructure unprecedented in scope.  In response to these trends, many commentators have argued that we are entering a new phase of science entirely\footnote{Michael Nielsen, \emph{Reinventing Discovery: The New Era of Networked Science}, Princeton 2011.}  and some have even questioned the role hypothesis driven investigation will continue to play in the era of ``big data.''\footnote{Editorial, ``Defining the scientific method,'' \emph{Nature Methods} \textbf{6}, 237 2009.}  There may or may not be lessons for us, but in such an era of turbulence and upheaval, it seems worth re-examining how the scientific revolution came about the first time around.  \\

The position that I will advocate in this essay is that the scientific revolution was indeed a period of dramatic intellectual change and that in a sense, what took place was a philosophical transformation.  But it is for reasons that I suspect most people will be wholly unfamiliar with.  Drawing upon the pioneering historical work of Frances Yates and Paolo Rossi, I will argue the following:
\begin{itemize}
\item 17th century Europe was a cauldron of ideas related to the art of memory.  
\item These ideas have their origin in ancient Greece and Rome, where the practice of memorization, using a visualization technique known as the ``method of places and images,'' was a foundational methodology for rhetoric.  
\item The classical art of memory was preserved through the Middle Ages largely in the monastic context, with St. Thomas Aquinas being one of the principle champions of the method of places and images.  Aquinas also introduced an important innovation by suggesting that these techniques could be used not only for the purpose of memory, but also as a concentration device for developing virtue and ethical behavior.
\item During the Renaissance, a critical transition took place wherein the mnemonic images of the classical art of memory were thought to not only serve the purpose of remembering, but to also represent the logical structure of nature itself.
\item In the 16th century, Giulio Camillo, a major intellectual figure in the Renaissance transformation of medieval memory, laid out an ambitious program to systematize all knowledge with a theater sized repository of standardized mnemonics.  
\item Somewhat independent of the classical art of memory, the 13th century Franciscan friar Ramon Lull proposed an alternate conceptualization of memory techniques in the context of combinatory wheels whose purpose was to eliminate logical contradictions from Christian theology.
\item At the arrival of the 17th century, there were several major schools of thought on memory and method in place.  The classical art of memory involving the method of places and images, the method of Ramon Lull and his combinatory wheels, and finally, the dialectic method of Petrus Ramus, with its emphasis on a logical structuring of knowledge and memorization by ordinary repetition.  
\item In the late 16th century and 17th centuries, Giordano Bruno and Gottfried Leibniz emerged as significant unifiers of these disparate and often conflicting intellectual traditions.  Bruno developed a hybrid mnemonic-Lullian method in which the method of places and images was used in conjunction with Lullian combinatory wheels, and Leibniz developed a hybrid mnemonic-dialectic method in which an encyclopedia was first developed to which each concept would then be assigned a corresponding mnemonic image.  
\item The fundamental concept in all of these developments is the general notion of ``method,'' as first proposed by Petrus Ramus.  The connotation of ``method'' is likely analogous to our modern word algorithm, and in the same way that the true meaning of algorithm only gains in substance when used in the context of computing technology, the word method gained its novel connotation in the context of the method of places and images, Lullian combinatory wheels, and the subsequent advances made by Bruno and Leibniz.
\item It was amidst these developments in methodological dialogue that Bacon, Descartes, and others proposed the scientific method, a method that aimed to simplify the various schools of memory, while preserving the core belief in a systematic approach to knowledge and secular outlook.  
\item Without the broad nature of developments in methodological thinking, without the awe inspiring feats one could perform with the art of memory, and the vision of all knowledge systematized in what we might reasonably call a computational framework, whether in Camillo's memory theater, or in Leibniz's universal calculus, the scientific method is unlikely to have taken root.  The art of memory provided an essential context and a vision of what would be possible with a systematic approach to knowledge.  The scientific method is effectively a distillation of the principles that were largely developed and made widely known in the context of the art of memory.  
\end{itemize}


This essay then is organized as follows.  I first give an overview of the several major periods of the \emph{ars memorativa}, or art of memory, starting from the Classical period in ancient Greece and Rome, and then progressing through the Middle Ages and the Renaissance.  The schools of thought that were in place by the end of the 16th century were the major conceptual battlegrounds from which the scientific method emerged.  I describe the major competing schools of thought of this time and provide several modern analogies that may help to shed some light into what the elite thinkers of this era were trying to accomplish.  I explain Petrus Ramus' notion of ``method" and how a perhaps more proper characterization of the 17th century would be ``methodological revolution'' rather than ``scientific revolution.''  I then return to the question ``what is the scientific method?'' and describe how Bacon's writings on the scientific method were one of the later arrivals onto the scene of methodological discourse and that prior interest in the various schools of thought on memory and method fundamentally enabled widespread adoption of the Baconinan perspective.  I will argue that what we have come to know as the scientific method is simply a small remnant of a profound set of ideas concerning the acquisition and development of knowledge, most of which have been forgotten.  Finally, I discuss Leibniz' project for the universal calculus and examine the acceleration of mathematical thought in light of 17th century philosophy.  So without further ado, let us turn our attention to ancient Greence, where as legend has it, the art of memory was born.  

\subsubsection*{The classical art of memory}
An apocryphal story attributes the origin of the art of memory to the lyric poet Simonides.  While attending a banquet where he was to recite a poem composed in the honor of a patron, Simonides is said to have been called outside by the request of two visitors.  While attending to the visitors, the banquet hall collapsed, killing everyone.  Unfortunately, the bodies and faces were so mutilated that the remains could not be identified, which would preclude family members from performing the proper funeral rites.  However, Simonides realized that the image of the banquet hall and those in attendance was well preserved in his mind, and by walking around the table in his mind's eye, he was able to recall the name of each person at the table and where they sat in relation to the others.  Thus, he was able to identify all of the remains so that the appropriate family members could be contacted. \\

From this experience, legend has it, Simonides concluded that the act of memorization could be enhanced by encoding objects and concepts into spatially organized visual images.  As Cicero would later write of Simonides in recounting the origin of the art of memory, 
\begin{quote}
{\small He inferred that persons desiring to train this faculty must select places and form mental images of the things they wish to remember and store those images in the places, so that the order of the places will preserve the order of the things, and the images of the things will denote the things themselves, and we shall employ the places and images respectively as a wax writing-tablet and the letters written on it.\footnote{Cicero, De oratore, II, lxxxvi, 351-354, quoted in Frances Yates, \emph{The Art of Memory}, London 1984, p. 3.}}
\end{quote}
He continues,
\begin{quote}
{\small It has been sagaciously discerned by Simonides or else discovered by some other person, that the most complete pictures are formed in our minds of the things that have been conveyed to them and imprinted on them by the senses, but that the keenest of all our senses is the sense of sight, and that consequently perceptions received by the ears or by reflexion can be most easily retained if they are also conveyed to our minds by the mediation of the eyes.\footnote{Ibid, p. 4.}}
\end{quote}

Where was this art form used?  The primary domain of application was in rhetoric.  Indeed, the major sources we have about the ancient art of memory come not from ancient Greece, but rather, from ancient Rome, where the method of places and images\footnote{I will use the phrases ``method of places and images,'' \emph{ars memorativa}, and ``method of loci'' more or less interchangeably in his essay.} was carried on as part of the fundamental training of students in rhetoric.  The three latin sources that largely inform our understanding of the ancient practice of this art are the \emph{ad Herrenium}, a Roman rhetoric textbook written by an anonymous teacher written around 86 BC, Cicero's \emph{de Oratore}, and Quintillian's \emph{Institutio oratorio}. \\

Practically, as the tradition was developed, the places that people would use were often actual locations or important buildings and other architectural creations-- one wonders how many facts have been encoded into mental images of the Parthenon, or later in ancient Rome, into mental images of the Colosseum.  Indeed, in our modern language, we see remnants of this fascinating cognitive use of architecture.  The phrase ``in the first place'' came into common usage when it was understood that the person speaking was using the method of places and images to encode the contents of a speech.  Consequently, the audience would have understood that the ``first place'' and the ``second place'' and so on, represented actual physical locations in the mind's eye of the speaker.  \\

Before moving on to later centuries and the evolution of the art of memory, let me mention a critical distinction made in the rhetorical tradition, primarily to illustrate the immense capacity for memorization via symbolic representation that these ancient orators would have developed using mnemonic techniques.  The \emph{ad Herrenium} and other texts distinguish between the ``memory for things'' and ``memory for words''.  The ``memory for things'' is the classic art of memory I have described, whereby an idea, or concept of some kind is represented using a striking image.  In then delivering the speech, although the order of topics is memorized, the precise words are delivered extemporaneously and not memorized a priori.  On the other hand, something that was also prescribed as a technical practice, and not for practical purpose, was the ``memory for words'' whereby a speech would be composed in its entirety and then memorized using the method of places and images \emph{word for word.}  This should truly astound a modern intellectual unfamiliar with these methods.  The fact that the memory for words was practiced at all, even purely for the purpose of strengthening the raw mental muscles of the natural memory, demonstrates what a phenomenal capacity these ancient rhetoricians must have developed, not only for memory, but in developing their inner sight and ability to visualize minute details with extraordinary clarity.  

\subsubsection*{The art of memory during the Middle Ages and the Rennaissance}
We see familiar prescriptions in a passage from the writings of Albertus Magnus, a 13th century Dominican Friar, 
\begin{quote}
{\small Those wishing to reminisce (i.e. wishing to do something more spiritual and intellectual than merely to remember) withdraw from the public light into obscure privacy: because in the public light the images of sensible things are scattered and their movement is confused.  In obscurity, however, they are unified and are moved in order.  This is why Tullius in the ars memorandi which he gives in the Second Rhetoric prescribes that we should imagine and seek out dark places having little light.  And because reminiscence requires many images, not one, he prescribes that we should figure to ourselves through many similitudes, and unite in figures, that which we wish to retain and remember.  For example, if we should wish to record what is brought against us in a law-suit, we should imagine some ram, with huge horns and testicles, coming towards us in the darkness.  The horns will bring to memory our adversaries, and the testicles the dispositions of the witness.\footnote{Albertus Magnus, \emph{Opera omnia}, ed. A. Borgnet, Paris, 1890, IX pp. 97, quoted in Yates, p. 82.}}  
\end{quote}

The same places and images of antiquity!  It is significant that during the Middle Ages, it was mistakenly thought that the \emph{ad Herrenium} was written by Cicero, or as he was often referred to, Tullius.  As I discussed above, Cicero's \emph{de Oratore} is one of the few artifacts we have of the art of memory during antiquity, but the \emph{ad Herrenium} was in fact written by an anonymous rhetoric teacher and not Tullius himself.  Still, during the Middle Ages, both works were taken together as the ``First and Second Rhetorics of Tullius,'' a mistaken attribution which would ultimately come to lend substantial momentum to a fundamental Medieval transformation.  The misattribution of the \emph{ad Herrenium} to Cicero proved to be critical, because in \emph{de Oratore}, he gives a great deal of attention to ethics and virtue as topical fodder for speeches.  And then in the ``Second Rhetoric'' he gave techniques for how those topics would be properly memorized.  This connection would ultimately lead to the Medieval shift of the art of memory from being a rhetorical technique to an ethical one.  In particular, it is Cicero's inclusion of memory as an integral part of the cardinal virtue of Prudence which proved vital to this evolutionary trajectory of the art of memory. \\

We can see this transition latent in St. Thomas Aquinas' four precepts for memory:
\begin{quote}
\small{Tullius (and another authority) says in his Rhetoric that memory is not only perfected from nature but also has much of art and industry; and there are four (points) through which man may profit for remembering well}
\begin{itemize}
\item[(1)] {\small The first of these is that he should assume some convenient similitude of things which he wishes to remember; these should not be too familiar, because we wonder more at unfamiliar things and the soul is more strongly and vehemently held to them; whence it is that we remember better things seen in childhood.  It is necessary in this way to invent similitudes and images because simple and spiritual intentions slip easily from the soul unless they are as it were linked to some corporeal similtudes, because human cognition is stronger in regard to the sensibilia.  Whence the memorative (power) is placed in the sensitive part of the soul.}
\item[(2)] {\small Secondly it is necessary that a man should place in a considered order those (things) which he wishes to remember, so that from one remembered (point) progress can easily be made to the next.  Whence the Philosopher\footnote{These quotations are references to Aristotle.} says in the book \emph{De Memoria} `some men can be seen to remember from places.  The cause of which is that they pass rapidly from one (step) to the next.'}
\item[(3)] {\small Thirdly, it is necessary that a man should dwell with solicitude on, and cleave with affection to, the things which he wishes to remember; because what is strongly impressed on the soul slips less easily away from it.  Whence Tullius says in his Rhetoric that `solicitude conserves complete figures of the simulacra.'}
\item[(4)] {\small Fourthly, it is necessary that we should meditate frequently on what we wish to remember.  Whence the Philosopher says in the book \emph{De Memoria} that `meditation preserves memory' because, as he says `custom is like nature.  Thence those things which we often think about we easily remember, proceeding from one to another as though in a natural order.'}\footnote{Quaestio XLIX , \emph{De singulis Prudentiae partibus}: articulus I, \emph{Utrummemoria sit pars Prudentiae} quoted in Yates, p. 67.} 
\end{itemize}
\end{quote}

We can take away a few points from Aquinas' precepts.  First of all, we recognize, of course, as in the excerpt from Magnus, the familiar rules for places and images.  And yet, whereas Magnus gives a classical example (from the \emph{ad Herrenium} in fact) with the ram and its horns as a mnemonic image for recalling the witness in a particular set of legal proceedings, we see in Aquinas writing the impact of a medieval piety, and the notion that ``spiritual intentions,'' rather than facts can be remembered and strengthened with this practice.\footnote{Of course, Magnus also hints at this idea in the preceding quotation: ``Those wishing to reminisce (i.e. wishing to do something more spiritual and intellectual than merely to remember) . . . ''} \\

There is much more to be said about Magnus and Aquinas, but having made the basic point that the medieval practice of the mnemonic arts took on a more spiritual rather than functional character, let me now to turn to a completely separate strain of thought during the Middle Ages which would come to feature quite significantly during the 17th century.  This line of thinking originated with Ramon Lull, a 13th century Franciscan friar, who developed an art of memory that proceeded along very different lines than the method of places and images, and which nevertheless had a great deal in common with the Thomist objective of strengthening one's spiritual capacity and virtuosity.  \\

Ramon Lull's technique was distinctly evangelical in its purpose.  The intended hope was that by arranging the fundamental concepts of Christianity on combinatory wheels, ultimately, one would be able to reduce the Christian corpus to its logical essence, and thus convince non-believers of the fundamental truth of the Christian gospel.  \\

Lull's combinatory wheels consisted of concentric circles, each of which was populated by symbols from the standard alphabet (see Figure 1).  Each symbol was used to represent a particular concept, for example, one set of nine concepts used by Lull consisted of ``Goodness, Greatness, Eternity, Power, Wisdom, Will, Virtue, Truth, Glory.''  Lull then created, or imagined, concentric wheels with these concepts on each wheel, as denoted by specific symbols or characters.  The wheels move independently, so by turning each wheel,  in the mind's eye of course, we arrive at a distinct set of concepts along a given column.  For example, if we have a setup with 3 concentric wheels, we might imagine that in one particular arrangement, a given column represents the concepts, ``Goodness, Will, and Eternity'' and by turning the second wheel by one notch we arrive at ``Goodness, Virtue, Eternity.''  It appears as through Lull's combinatory wheels served two purposes.  They provided a framework in which one was to structure core Christian values-- presumably, some amount of effort went into the specific choice of concepts and symbols-- and second, they provided a practical technique for a practitioner to first enumerate and then focus on a particular grouping of concepts.  What Lull's intent may have been in believing that this process would ultimately allow one to reduce Christian thought to its logical essentials and ultimately, to free it of all logical contradictions, was that it provided a systematic method for contemplating all possible groupings of different concepts from their respective categories.  Then over the course of extended contemplation and discussion, one would engage in a long-term process of re-evaluating and evolving the scriptural foundations.  Lull's combinatory wheels provided a systematic set of procedures to go about this ambitious process.  Along the way, it ensured that each combination would receive careful attention, and hopefully, would draw attention to any possible inherent logical conflicts at the basis of Christian thought.  \\

\begin{figure}\label{fig:lull}
\begin{center}
\includegraphics[width=.6\textwidth]{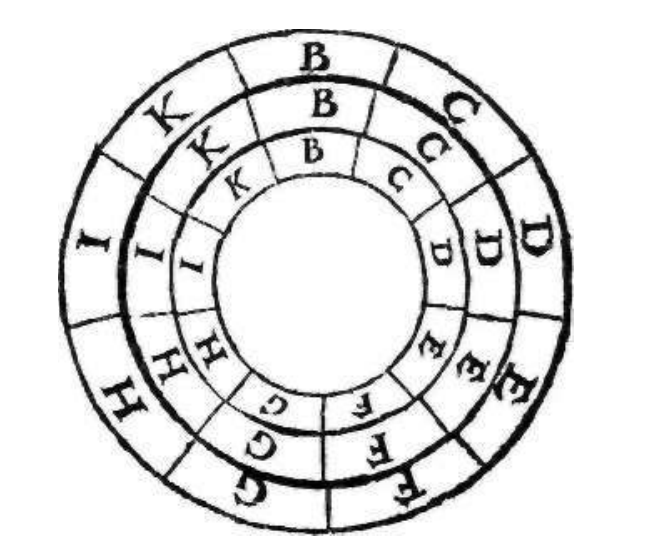}
\caption{{\small Lullian combinatory wheel, from \emph{Art brevis}.\protect \footnotemark}}
\end{center}
\end{figure}

\footnotetext{Yates, Fig. 6, p. 183.  Reproduced with permission.}
What is remarkable about Lull's revolving wheels, is how lacking they are in any kind of aid to memory in any spatial or visual sense.\footnote{Lull did develop another method which is more reminiscent of the classical art, in which different branches of knowledge were to be organized on trees.  Unlike the combinatory wheels, these tree-like structures would have enabled the memory in a more direct way.  Still, I believe that Yates and Rossi are both correct in including Lull's work in the lineage of the \emph{ars memorativa}.  First of all, even though the combinatory wheels may not have enabled the memory, it clearly prioritized memory in a significant manner-- one way or another, an aspiring practitioner would have to develop his or her memory substantially to put these techniques into use.  If the goal was simply to enumerate all possible combinations of a particular group of concepts, why not simply do so on slips of paper, parchment, or cloth?  It is clear that Lull believed that memory was essential to the long-term vision of reducing Christian thought to its logical essence, a process that would be enabled by having practitioners hold the different concepts in their mind in a focused, meditative manner.  But more importantly, in understanding the 17th century and the emergence of the scientific method, Lull's combinatory wheels would feature significantly in later efforts that aimed to bridge competing schools of thought on memory.}  Whereas the classical tradition seems to provide its practitioner with a powerful tool to boost the memory, Ramon Lull's method seems to demand, from the outset, an inner sight of penetrating clarity.  As a simple exercise, I encourage the reader to visualize nine letters arranged in a circle, and then try to imagine rotating the positions of the letters one by one.  It is not easy!  Now imagine that these letters are to be equally spaced on concentric circles, with each letter representing a concept that must be remembered on its own.  Only after holding all this information in one's mind is the practitioner in a position to begin to rotate the position of one of the wheels!  If Ramon Lull alone had the ability to perform this feat, it would be noteworthy.  If, as the historical record seems to indicate, there was an entire school of thought devoted to his practice, it is truly a testament to the neural plasticity of the adult brain, and its capacity to train new and unusual cognitive abilities with enormous amounts practice and concentration.  Unlike the classical art, the method of Ramon Lull does not seem to take advantage of an innate capacity to remember places and images-- rather it places a huge burden on our raw memory and visual ability and aims to develop them both in a single-pointed fury of devotion.  \\

Yet another important transition took place in the philosophy and practice of the art of memory during the Rennaissance.  Whereas mnemonics had previously been thought of as being tools for remembering, and in some cases, for instilling and intensifying religious devotion and developing certain virtues, a new notion emerged during the Renaissance.  In this new school of thought, it was put forth that relationships between the objects of mnemonic devices could perhaps be encoded into the relationships of the mnemonics themselves, and thereby give rise to the ability to understand the natural world by the strategic use of mnemonics.  \\


The most significant figure in this tradition was Giulio Camillo, a 16th century philosopher and scholar, who devoted the majority of his life to a project so fantastic and magnificent in scope that the entirety of Europe was abuzz in excitement of the consequences.  The project was to build a theater sized repository of mnemonic figurines, physical statues which corresponded to the classical method of places and images, but which would have been standardized through great effort, and which would represent the near totality of contemporary thought.  Much of Camillo's theater would have been devoted to religious and astrological concepts. The theater, which was never built, consisted of seven grades, or steps (representing the seven planets), each of which was devoted to a particular branch of knowledge.  The entrant, who would have been well versed and practiced in the classical art of memory, would then be confronted with cabinets containing mnemonic images representing some specific concept, fact, or idea.  In addition, there were numerous drawers which were filled with papers of speeches from Cicero relating to the topical matters of the images depicted in the surrounding areas.  Viglius Zuichemus, a Dutch statesman, described the theater and his meeting with Camillo in a letter to his contemporary, Erasmus:

\begin{quote}
{\small The work is of wood, marked with many images, and full of little boxes; there are various orders and grades in it.  He gives a place to each individual figure and ornament, and he showed me such a mass of papers that, though I always heard that Cicero was the fountain of richest eloquence, scarcely would I have thought that one author could contain so much or that so many volumes could be pieced together out of his writings.  I wrote to you before the name of the author who is called Julius Camillus.  He stammers badly and speaks Latin with difficulty, excusing himself with the pretext that through continually using his pen he has nearly lost the use of speech.  He is said however to be good in the vernacular which he has taught at some time in Bologna.  When I asked him concerning the meaning of the work, its plan and results--  speaking religiously and as though stupefied by the miraculousness of the thing--  he threw before me some papers, and recited them so that he expressed the numbers, clauses, and the artifices of the Italian style, yet slightly unevenly because of the impediment in his speech.  The King is said to be urging that he should return to France with the magnificent work.  But since the King wished that all the writing should be translated to French, for which he had tried an interpreter and scribe, he said that he thought he would defer his journey rather than exhibit an imperfect work.  He calls this theater of his by many names, saying now that it is a built or constructed mind and soul, and now that it is a windowed one.  He pretends that all things that the human mind can conceive and which we cannot see with the corporeal eye, after being collected together by diligent meditation may be expressed by certain corporeal signs in such a way that the beholder may at once perceive with his eyes everything that is otherwise hidden in the depths of the human mind.  And it is because of this corporeal looking that he calls it a theater.\footnote{Erasmus, \emph{Epistolae}, ed. P. S. Allen et al., IX, pp. 29-30 quoted in Yates, p. 132.}}
\end{quote}

Camillo's theater represents a third and distinct phase of the art of memory.  What began in antiquity as a functional practice for memorizing the contents of speeches in the rhetorical tradition, then took on a spiritual character during the Middle Ages in the hands of St. Thomas Aquinas and the Dominican tradition.  In a sense, Camillo's mnemonic theater fuses the classical art of memory with part of the intent of Ramon Lull's quite distinct practice of religious concepts encoded into combinatory wheels.  Camillo's theater, while still in the tradition of the method of place and images, represents a vision of knowledge systematized to a heroic degree, not solely for the purpose of memory, but for understanding the structure of knowledge itself and for accurately reflecting the basic principles of the cosmos in its mnemonic organization.   

\subsubsection*{The 17th century and a conceptual reinvention}
Entering the 17th century, we are confronted with the following raw materials in the world of method and memorization.  From ancient Greece and ancient Rome, via the rhetorical tradition and through the work of Dominican order comes the tradition of the \emph{ars memorativa} and the systematic and creative use of mnemonics for the purpose of memory, and in a closely related capacity, for ethical development.  On the other hand, from Ramon Lull, who we recall as early as the 13th century was developing his own system of symbolic representation of religious concepts, we have the notion of an intensive practice of memory, but without the associated system of images.  From Giulio Camillo, we have the Renaissance transformation of the classical art, with its rules for images and places well preserved, but now with the additional notion that mnemonic images could also be used to represent the structure of knowledge and of Nature itself.  \\

And of course, memory through ``mere'' repetition is an age old notion that has no name to which we can attribute a founder, although it was more precisely formulated by a man who is critical in understanding the 17th century and the origin of the scientific method, Petrus Ramus.  Ramus was a 16th century educational reformer who was particularly concerned with re-examining the ways in which subjects were to be memorized.  In particular, Ramus made the dramatic, iconoclastic move of eliminating memory as part of rhetoric.  With this, he was able to distance himself from the mnemonic practice of places and images, which he replaced with a new method, called the ``dialectic method.''  In this method, a subject is first structured in a logical manner by proceeding from the most general concepts to the most specific.  The content is then memorized by the standard practice of repetition.  \\

Part of Ramus' motivation as a reformer related to a specific set of religious objections to the classical art of memory which stem from the Old Testament:
\begin{quote}
{\small Take ye therefore good heed unto yourselves; for ye saw no manner of similitude on the day that the Lord spoke unto you in Horeb out of the midst of the fire: Lest ye corrupt yourselves, and make you a graven image, the similitude of any figure the likeness of male or female . . . And lest thou lift up thine eyes, unto heaven, and when thou seest the sun, and the moon, and the stars, even all the host of heaven, shouldst be driven to worship them . . . \footnote{P. Ramus, \emph{De religione Christiana}, cd. of Frankfort, 1577, pp. 114-15 quoted in Yates, p. 236.}}
\end{quote}
According to Yates, this prohibition of graven images, taken from the fourth chapter of Deuteronomy, was interpreted by Ramus as applying to the classical art of memory as well.  The rhetorical tradition quite actively advocated the use of lewd and grotesque images by which to excite the imagination and empower the memory, which to Ramus, was tantamount to a systematic technique for polluting one's mind.  \\

It is not too difficult to sympathize with Ramus when considering some of the writings of his predecessors.  Pietro de Ravenna, a 15th century jurist, ardent self-promoter, and evangelist of the method of places and images wrote the following as a suggested set of techniques for inventing more effective mnemonics:
\begin{quote}
{\small I usually fill my memory-palaces with the images of beautiful women, which excite my memory . . . and believe me: when I use beautiful women as memory images, I find it much easier to arrange and repeat the notions which I have entrusted to those places.  You now have a most useful secret of artificial memory, a secret which I have (through modesty) long remained silent about: if you wish to remember quickly, dispose the images of the most beautiful virgins into memory places; the memory is marvelously excited by images of women. . . This precept is useless to those who dislike women and they will find it very difficult to gather the fruits of this art.\footnote{Pietro de Ravenna, \emph{Phoenix sue artificiosa memoria}, quoted in Paolo Rossi, \emph{Logic and the Art of Memory}, Chicago 2000, p. 22.}}
\end{quote}

It is worth noting that there is a purely secular counterpart to the Ramist objection, not on the grounds of spiritual pollution, but rather, conceptual interference.  As I have described, and as an any reader who chooses to experiment with these techniques will see for him or herself, the classical method of places and images gains its strength by taking advantage of vivid conceptual associations, which almost always have very little in common with the specific facts or ideas that are being remembered.  It is necessarily the case then, that by using these techniques extensively, one is creating a vast array of associations that are completely arbitrary and have no resemblance to the logical structure of content being memorized.  \\

In attempting to circumvent these stray associations, whether on religious or secular grounds, Ramus introduced a concept that will prove to be critical in understanding the 17th century and the scientific revolution.  Ramus was the first thinker to popularize the word ``method.''\footnote{Yates, p. 369.}   In the modern world, we understand the connotation of ``method'' as referring to an orderly, procedural practice, but in the 16th century when Ramus began to popularize this word, recall that he was advocating a specific type of method-- the dialectic method-- which stood in contrast to the ``mnemonic method,'' i.e., the classical art of memory, and the ``method of Ramon Lull,'' consisting of combinatory wheels with associated symbols.  \\

To understand the specific connotation of this word as it would have been understood by Ramus and his contemporaries, I believe we should examine the modern word \emph{algorithm}.  The Oxford-English Dictionary defines an algorithm to be,

\begin{quote}
{\small A procedure or set of rules used in calculation and in problem solving; a precisely defined set of mathematical or logical operators for the performance of a particular task.\footnote{Algorithm [Def. 2]. (n.d.). Oxford English Dictionary Online. In Oxford English Dictionary. Retrieved May 16, 2013, from http://www.oed.com.}}
\end{quote}

I think many would agree that something is lacking in this definition.  Specifically, the word algorithm is a fairly recent word, and yet this definition describes a notion that has existed for thousands of years.  In particular, Euclid's algorithm for finding the greatest common divisor of two numbers dates back as far as 300 BC.  What seems to be critically missing from the definition, is the additional \emph{connotation} the word algorithm gains when used in reference to modern computing technology.  Certainly, young children can learn algorithms, say for performing long division, well before they are exposed to computer programming, and teaching them the word algorithm in this context would add very little to their conceptual maturity.  However, the word itself gains substantial depth with the additional notion that a computing device can perform a set of instructions thousands, millions, and billions of times, with a precision and accuracy that no human could otherwise accomplish.  This additional connotation, which is experienced through interacting with computers, through film and television, and which is not captured in the purely dictionary definition, is critical, and I believe we can re-examine the word ``method,'' as used by Ramus, in a similar light.  \\

While Ramus may have objected to the classical use of images and places, his own dialectical method-- one particular kind of ``method''-- certainly profited from the connotation that the art of memory carried with it, namely as a systematic procedure for the memorization of knowledge.  But furthermore, Ramus' dialectic method had also something in common with the ``method of Ramon Lull,'' as both methods aimed to distill a particular knowledge base, in Lull's case the Christian doctrine, to its logical essence.  Indeed, Lullian combinatory wheels may have played an essential role in lending a practical set of associations to the word ``method,'' in the same way that computing technology is essential in understanding the word algorithm in modern times.  Recall again, the standard Lullian setup, say for example,  consisting of 3 concentric circles with 9 Greek symbols designating the concepts Goodness, Greatness, Eternity, Power, Wisdom, Will, Virtue and Glory.  Lull's method for enumerating all $9 \choose 3$ $= 84$ distinct combinations provides a systematic procedure for ensuring that in the process of contemplating the implications and interrelation of each concept with each other, no stone would be left unturned.  \\

We have seen thus far that the notion of ``method'' unified the diverse strains of thought related to memory, and that in an important way, the pure act of memorization was secondary to the notion of method as a systematic procedure for acquiring knowledge and investigating Nature.  Amid these turbulent conceptual battles being fought on the grounds of the mnemonic, dialectic, and Lullian methods, with disagreements about the very foundations of method,  there were two great mollifiers and unifiers that played a crucial role in the emergence of the mathematical and scientific methods, Giordano Bruno and Gottfried Leibniz.  \\

Bruno and Leibniz were both thoroughly versed in all branches of methodological thinking, from the classical method of places and images, to the method of Ramon Lull, to the dialectic method of Ramus.  Born four years after the death of Giulio Camillo, Bruno was trained in a Dominican convent in Naples, where he would have certainly been exposed to the method of places and images from the \emph{ad Herrenium}, and to the work of his predecessor, St. Thomas Aquinas.  Bruno's first treatise on memory, \emph{De umbris idearum} was published in 1582, and was dedicated to the King of France:
\begin{quote}
{\small I gained such a name that the King Henri III summoned me one day and asked me whether the memory which I had and which I taught was a natural memory or obtained by magic art; I proved to him that it was not obtained by magic art but by science.  After that I printed a book on memory entitled \emph{De umbris idearum} which I dedicated to His Majesty, whereupon he made me an endowed reader.\footnote{\emph{Documenti delta vita di G.B.}, ed. V. Spampanato, Florence, 1933 pp. 84-85, quoted in Yates, p. 200.}}
\end{quote}

Bruno's early work had much in common with the tradition of Camillo, with a significant emphasis on the use of places and images for astrological purposes.  Having been brought up in a world in which Camillo's influence would have been strongly felt and talk of his magnificent theater quite active, and in addition, belonging to the same Dominican order as St. Thomas Aquinas, Bruno in a sense was heir to the most significant historical developments stemming from the classical art of memory.  But his historical position is significantly colored by exposure to that other thread of the art of memory, the method of Ramon Lull.  Indeed, Paris, not far from Bruno's hometown of Naples, was the epicenter of 16th century Lullism, and it was in combining these two traditions that Bruno would make his mark.  \\

Bruno took the bold move of unifying both traditions by starting with Lullian combinatory wheels, but using classical mnemonic images instead of the standard alphabet for representing concepts on each locus.  Thus, he replaces the architectural component of the method of places and images by a Lullian combinatory wheel.  Furthermore, in contrast to Ramon Lull, Bruno's objectives were not strictly religious in nature.  Indeed, he had a great deal in common with Giulio Camillo, and we can see the extent of this vision in the shocking size and complexity of Bruno's wheels.  In a particular example extracted from one of his treatises, we see detailed a combinatory wheel with 30 divisions, each of which is further divided into 5 subdivisions, giving 150 divisions in total.  The lists that he includes in the book are sets of 150 images each, which are to be ordered on the wheels in a Lullian fashion.\footnote{Yates pp. 217-223 and Rossi pp. 87-88}  In the remainder of this essay, I will refer to Bruno's system as the hybrid mnemonic-Lullian method.  \\

The complexity of such a system is truly appalling.  I invite the reader, if he or she has not already done so, to attempt the simple exercise I outlined earlier of visualizing letters arranged in a circle.  Having gotten to the point of visualizing this structure clearly, now attempt to rotate the position of each letter.  If this exercise seems demanding, imagine what kind of mental facility would be required to manipulate the hundreds of detailed images populating a Brunian mnemonic wheel!  As I stated of Ramon Lull, if Bruno alone could visualize an object of such stunning complexity, it would have been a noteworthy accomplishment, if only for demonstrating the human mind's capacity for training unusual cognitive abilities.  \\

It may surprise readers that Gottfried Leibniz, who most know as being the co-inventor of the infinitesimal calculus, was one of the foremost figures in methodological innovation.  Indeed, his role during the 17th century paralleled that of Bruno, and whereas Bruno attempted to unify the Lullian and mnemonic traditions, Leibniz' primary effort, indeed what we may reasonably describe as his overarching vision, was an effort to unify the mnemonic method with the dialectic method of Petrus Ramus.\footnote{I am diverging from Yates and Rossi in connecting Leibniz with Ramus.  Strictly speaking, Leibniz' work on the universal calculus should be properly situated, in addition to its obvious heritage stemming from the art of memory, in the context of the encyclopedic tradition.  But particularly in comparing Leibniz' efforts to those of Giordano Bruno, I find it most illustrative to refer to Brunian mnemotechnics as a hybrid mnemonic-Lullian method and Leibniz' method as a hybrid mnemonic-dialectic method.  I certainly will have no objection if future scholars think to overturn this distinction-- my primary interests in this article do not rest on this particular choice of nomenclature.}  Certainly Leibniz was intimately familiar with the Lullian tradition and also of Bruno's attempts at unification.  The ``universal calculus'' was Leibniz' primary aim and it borrowed from both the dialectical and mnemonic traditions in the following manner. First, an encyclopedia was to be constructed covering the entire domain of human thought, from science, to religion, to law.  In the spirit of the dialectic method, the encyclopedia would be carefully assembled so as to reflect the natural logical structure of each discipline.\footnote{An example might help to illustrate Leibniz' views on what a logical structuring of knowledge entailed.  One of the primary intermediate goals Leibniz hoped would ultimately lead to a universal calculus was to determine all possible subjects of a given predicate, and conversely, given a subject, to determine all possible predicates.  To bridge this encyclopedic, or dialectic, objective with the notion of a symbolic calculus, Leibniz proposed that after fixing a pre-determined ordering of possible predicates, one could associate to each predicate the corresponding prime number (with respect to the fixed ordering).  Then, for a subject which satisfied some number of predicates, one could uniquely associate a natural number by taking the product of the corresponding prime numbers.  See, for example, Rossi pp. 178-179.}  Finally, drawing from the mnemonic method, symbols would be constructed from each core concept, from which, Leibniz hoped, a \emph{universal calculus} would emerge in which logical contradictions could be eliminated from the entirety of human thought and in which all questions, whether legal, scientific, or religious could be answered by \emph{computing the answer} via manipulation of the associated symbolic infrastructure.  As we see in the following passage, it is clear that Leibniz viewed this project as requiring permanent, ongoing evolution, progressing hand in hand with the development of scientific knowledge:
\begin{quote}
{\small Although this language (the universal calculus) depends on true philosophy, it does not depend on its perfection.  Let me just say this: this language can be constructed despite the fact that philosophy is not perfect.  The language will develop as scientific knowledge develops.  While we are waiting, it will be a miraculous aid: to help us understand what we already know, and to describe what we do not know, and help us to find the means to obtain it, but above all it will help us to eliminate and extinguish the controversial arguments which depend on reasons, because once we have realized this language, calculating and reasoning will be the same thing.\footnote{Couturat, \emph{Opuscules et fragments in\'{e}dits de Leibniz}, pp. 27-28, quoted in Rossi, pp. 174-175.}}
\end{quote} 

Leibinz seemed to be keenly aware of the importance of proper notation,\footnote{Stephen Wolfram discusses Leibniz' preoccupation with notation in the beautiful essay, ``Dropping in on Gottfried Leibniz,'' \url{http://blog.stephenwolfram.com/2013/05/dropping-in-on-gottfried-leibniz/}} and like Ramus, he also seemed to show concern for the potential for conceptual interference that the mnemonic method necessarily entailed.  But whereas Ramus chose to abandon the notion of symbolic representation entirely, Leibinz continued to express hope that one could find a set of symbols that would both enable the memory and serve as a proper foundation for what we might call a  ``calculus of concepts.''  Indeed, Leibniz hoped that by borrowing geometric elements from the Egyptian alphabet, or pictorial elements from the Chinese alphabet, he would arrive at a set of symbols which would precisely satisfy these competing constraints.\footnote{Rossi pp. 179-180 and Yates p. 381.}

\subsubsection*{The emergence of the scientific method}
We are now in a position to tackle the question I posed at the beginning of this essay, ``what is the scientific method?''  As I stated in the introduction, the standard answer is not entirely inaccurate.  As the Oxford English Dictionary states, the scientific method could certainly be defined as ``a method of procedure that has characterized natural science since the 17th century, consisting in systematic observation, measurement, and experiment, and the formulation, testing, and modification of hypotheses.''  And Francis Bacon was certainly among the first to articulate these principles. \\

What may surprise readers is that Bacon was not only aware of the art of memory, but also quite well versed in all aspects of methodological thinking.  Indeed, in \emph{De augmentis scientiarum}, we see a very familiar description of the method of place and images:
\begin{quote}
{\small Emblems bring down intellectual to sensible things; for what is sensible always strikes the memory stronger, and sooner impresses itself than the intellectual . . . And therefore it is easier to retain the image of a sportsman hunting the hare, of an apothecary ranging his boxes, an orator making a speech, a boy repeating verses, or a player acting his parts, than the corresponding notions of invention, disposition, elocution, memory, action.\footnote{\emph{De augmentis scientiarum}, V, ed. Spedding, I, p. 649, quoted in Yates, p. 371.}}
\end{quote}
But as a historical figure, Bacon perhaps most closely resembled Petrus Ramus as an educational theorist, and in \emph{The Advancement of Learning}, the art of memory is featured as one of the arts and sciences in need of reform.  In particular, Bacon's primary objection was that the art of memory could be used for unnecessarily acquiring massive amounts of information simply for the purposes of impressing people, and that this kind of effort would be more usefully directed at advancing the arts and sciences.  \\

Renes Descartes, another central figure in the emergence of the scientific method, also took on a Baconian interest in reforming the art of memory.  Perhaps coming more prominently from the tradition of natural philosophy and mathematics, Descartes was interested in using the art of memory for understanding causality.  In this fascinating passage, we see Descartes drawing a fundamental connection between the mnemonic method and reductionism:
\begin{quote}
{\small On reading through Schenkel's\footnote{Schenkel was a 17th century memory theorist.} profitable trifles (in the book \emph{De arte memoria}) I thought of an easy way of making myself master of all that I discovered through the imagination.  This would be done through the reduction of things to their causes.  Since all can be reduced to one it is obviously not necessary to remember all the sciences.  When one understands the causes all vanished images can be easily found again in the brain through the impression of the cause.  This is the true art of memory and it is plain contrary to (Schenkel's) nebulous notions.  Not that his (art) is without effect, but it occupies the whole space with too many things and not in the right order.  The right order is that the images should be formed in dependence on one another.  He (Schenkel) omits this which is the key to the whole mystery.  \\

I have thought of another way; that out of unconnected images should be composed new images common to them all, or that one image should be made which should have reference not only to the one nearest to it but to them all-- so that the fifth should refer to the first through a spear thrown to the ground, the middle one through a ladder on which they descend, the second one through an arrow thrown at it, and similarly the third should be connected in some way real or fictitious.\footnote{Descartes, \emph{Cogitationes privatae} (1619-1621); in \emph{Oeuvres}, ed. Adam and Tannery, X, p. 230, quoted in Yates, pp. 373-374.}}
\end{quote}

As has been the common theme throughout this essay, we see that the art of memory, sagaciously discerned by Simonides, or else discovered by some other person, has once again captivated the imagination of another great figure of the Age of Reason.  With Bacon and Descartes as my final historical examples, I hope to have convinced the reader at the very least, that the classical art of memory and its many descendants played a critical role in Western intellectual history.\footnote{As I stated in the introduction, the historical basis for this essay is largely from Frances Yates' \emph{The Art of Memory} and Paolo Rossi's \emph{Logic and the Art of Memory}.  Therefore, one primary contribution of this work is to draw attention to a fascinating aspect of scientific history that is perhaps not well known outside of a few academic circles.  Certainly, those interested in pursuing this line of thought should start with Yates and Rossi's work, and also Caruther's \emph{The Memory Book} and the references contained therein.  So as to interest a broad audience, and in particular, those in scientific fields, I have condensed a tremendous amount of history into these few pages and skipped over quite a few important figures along the way.  One could easily devote an entire lifetime to the intellectual history of the art of memory, and I hope that this essay will stimulate interest in professional and amateur historians alike.  It should be clear to the reader that I have chosen to focus my attention in this essay entirely on developments that took place in Europe.  I do not mean to imply that significant efforts in the art of memory did not take place elsewhere.  Indeed, there is an incredibly rich tradition of the memory arts in India and it does seem to be of historical importance to determine whether in the many scholastic traditions that have emerged across the world, if there have been developments that paralleled those that unfolded in Western Europe, particularly with regard to mnemonic techniques and methods involving symbolic representation, and not simply memorization through repetition.}  Let us now return to the origin purpose of this essay and re-examine the question ``what is the scientific method?'' \\

Most scientists will likely agree with the statement that good scientific practice largely involves perseverance, determination, curiosity, systematic thinking, patience, and experience.  And I suspect that most will agree that many of these qualities are the critical determinants of success in other, non-scientific fields as well.  Conversely, bad science can be characterized by efforts lacking in systematization, lacking in curiosity, impatience, and a lack of experience.  And I think most would agree that these qualities are learned, as I stated early on in this essay, by simply participating in the frontiers of research under the guidance of more experienced researchers.  \\

The observation that scientific maturity is largely communicated via implicit, cultural knowledge suggests that the development of the scientific method was not a discrete event.  As I stated in the introduction, we can certainly point to numerous examples of principled scientific reasoning well before the 17th century.  Indeed it would be difficult to imagine that hypothesis-driven investigation in some limited form or another was not always part of human society.  To properly re-contextualize the original question, therefore, it seems more appropriate to ask ``how and why did the adoption of the scientific method accelerate during the 17th century?''  \\

There is no doubt that the more commonly attributed elements of the scientific revolution were significant-- a vision of the natural sciences built upon hypothesis-driven investigation, the establishment of a scientific journal, and the critical support of wealthy patrons, etc.  But the question remains as to why these factors gained traction during this time period when they had not before.  Without a critical mass of scientific accomplishments, these principles, which seem elementary and self-evident to us, would have been rather abstract to individuals of that era.  What would have been the motivation to adopt a worldview which amounts to a highly philosophical, meta-scientific position towards the systematic pursuit of knowledge? \\

It is specifically in this context, that is, in providing a critical momentum and inspiring vision to motivate the widespread adoption of a scientific viewpoint that I believe the art of memory to have played an instrumental role.  In particular, the cultural and institutional circumstances of 17th century Europe provide an answer to the question of how philosophical argumentation concerning the principles of reason and using rational thought to understand Nature could have taken root in a world without the academic infrastructure that we have today.  What I believe to have been a critical enabler of scientific thought was that the two medieval schools of memory which would figure prominently in the 17th century, the classical art as preserved by St. Thomas Aquinas and the Dominican order, and the method of Ramon Lull preserved by the Franciscans, would have been well known throughout Europe, given the geographic flexibility of the friars.  Thus, not only would the most recent developments be intelligible to newcomers, for instance Bruno's hybrid mnemonic-Lulian method, or Leibniz' hybrid mnemonic-dialectic method, but in addition, there was a specific set of techniques that one could put into practice in order to understand the implications of the larger philosophical principles and world views being advocated.  \\

Without a concrete set of actions for listeners to engage with, it seems highly unlikely that the Baconian perspective would have resonated with a sufficiently large group of people to precipitate institutional or cultural change.  Imagine being a moderately or even highly educated member of 17th century European society.  You attend a lecture by Bacon on the virtues of reason and the notion that humanity is best served by the most educated and talented minds directing their efforts toward the natural sciences.  Even if you were inspired by these words, what is the next step?  What does one \emph{do} next?  In a world in which survival itself was a more serious undertaking, and in a continent torn by political instability, and in which a major plague took the lives of 15\% of the London population, no matter how inspiring the rhetoric, it seems more likely that a person's attention would be strongly pulled back to practical realities.  \\

On the other hand not only did the mnemonic method, the dialectic method, and the method of Ramon Lull all carry with them analogous rhetorical potency, they slowly began to be directed towards the study of the natural world and the development of the arts and sciences.  In addition, all of these methods were concrete practices that would have been universally understood, and which provided an inspiring vision of what could be accomplished with a systematic, rational approach to the pursuit of knowledge.  If one was familiar with the mnemonic method, and was then exposed to Leibniz' notion of the universal calculus, one could, in principle, begin immediately experimenting with these new ideas, or at the very least contemplate them and discuss them with others.  One could fall asleep at night, slowly pacing through thousands of memory loci, imagining that rather than being used for memorizing the words of Cicero, these same loci could be mathematical symbols which encoded a calculus of the natural sciences.  These \emph{internal experiences} would have been powerful, and thus, I believe that the true agent of philosophical change in the 17th century was the notion of ``method,'' and in particular, the descendants of the classical art of memory.  \\

If these powerful mental abilities were directed solely at rhetoric, we can imagine that they would have no bearing on future scientific institutions.  But the historical record indicates otherwise.  Indeed, over the course of nearly two millennia, we see that the art of memory went through multiple transformations and became increasingly directed at general knowledge and the unearthing of natural principles. And in the eyes of some of the greatest minds of the Western intellectual tradition, the art of memory was viewed as something substantially more general and versatile than merely a tool for memorization.  Indeed, nearly three centuries after the invention of the printing press, we see that the art of memory continued to be investigated, developed and improved upon, in the diverse incarnations of the mnemonic method, the dialectic method, the method of Ramon Lull, and the mnemonic-Lullian method of Giordano Bruno, and the mnemnonic-dialectic method of Gottfried Leibniz.  And not only were those pioneers traditionally associated with the rise of the scientific method, Francis Bacon and Renes Descartes, well-versed practitioners of these ``other'' kinds of method, but their outlook and vision for the continued development of intellectual institutions were formulated in reference to the memory arts. \\

One way to conceptualize the intellectual milieu of this time period is to recognize that the phrase ``scientific method'' is in fact a compound construction consisting of the two words ``scientific'' and ``method.''  Just as ``organic chemistry'' is a specific kind of chemistry as distinguished from ``physical chemistry'' and just as ``quantum mechanics'' is a specific kind of mechanics as distinguished from ``classical mechanics,'' the ``scientific method'' is a specific kind of method which at one point in time was distinct from other kinds of method such as the dialectic method, the method of Ramon Lull, the mnemonic method, and so on.  As I have argued, scientific thinking has always been part of human society-- what took place during the 17th century was that scientific thinking became fused with methodological thinking.  In other words, it was the art of memory that created the the ``method'' in ``scientific method.''  \\

In a sense, the principles articulated by Bacon, Descartes, and others, were something of the aftermath of what should perhaps be called the ``methodological revolution'' rather than the ``scientific revolution."  The Baconian school was in a position to survey the massive intellectual transformation that was created by the diverse manifestations of the memory arts, and these writings serve to distill the basic qualities embodied by the widespread practices of ``method,'' and to focus these efforts even more strongly in the direction of the natural sciences.  But without the prior history of the mnemonic method and its many descendants, this perspective may not have had nearly the same impact. 

\subsubsection*{The art of memory and the growth of the mathematical method}
I hope to have convinced the reader of the significant impact the art of memory has had on Western intellectual history.  It is a practice that has been the foundation of so many institutions and cultural movements, from Greek and Roman rhetorical students and statesman in antiquity, to the Franciscan and Dominican friars in the Middle Ages, to the natural philosophers and scientific trailblazers of the 17th century.  For an idea that has so profoundly sustained and nurtured Western civilization, it made a truly graceful exit.  \\

But perhaps it is an opportune moment in history to reconsider its value, and in the remainder of this essay, I will examine the role 17th century discourse played in the growth of mathematical thought.  In particular, I will examine Gottfried Leibniz' project to construct a \emph{universal calculus} which would resolve logical conflict in all areas of knowledge.  My primary aim in this section is to try to understand details of Leibniz' research agenda in greater depth, and in particular, to attempt to articulate what to modern mathematical scientists appears to be a rather peculiar set of beliefs about mnemonics and their potential for conceptual abstraction.  Along the way-- in some sense, attempting to view the world as Leibniz or one of his followers would-- I will argue the following:
\begin{itemize}
\item Mnemonics and mathematics both share the core property of being symbolic representations of concepts. 
\item In an era where mathematical thought was substantially less developed, and in particular, where there were far fewer applications of mathematics to the natural sciences, the difference between mnemonics and mathematics may have been viewed as being rather small.
\item Leibniz, in particular, may have viewed his work on the infinitesimal calculus as simply a ``toy'' problem in the much larger vision of the universal calculus, in which mnemonics would have played a critical role in creating a symbolic representation of all human knowledge.  That is, mnemonics were viewed as being a potential basis for physical theories, along side more traditional mathematics.  
\item Conversely, the motivation to pursue mathematical theories for physical phenomenon may have gained additional momentum via its affiliation with mnemonics, and in particular, the mnemonic theater of Giulio Camillo, Bruno's hybrid mnemonic-Lullian method, and Leibniz's hybrid mnemonic-dialectic method.  In other words, even though the mnemonic method ultimately did not give rise to physical models, there was a widespread vision and tremendous confidence that it would someday.  This vision may have helped to accelerate interest in pursuing mathematical approaches to model physical phenomena.  
\item The 17th century vision of mnemonics as being closely related to mathematics may provide us with a novel attack on the foundational questions of ``what is mathematics?'' or ``why can mathematics be used to model natural phenomena?''  
\end{itemize}

We live in an era where mathematics and the physical sciences have advanced to such an extent that we are able to make conceptual distinctions that would seem quite foreign to even the most sophisticated minds of a few centuries ago.  Mathematical researchers categorize analysis, algebra, and topology as clearly distinct topics from which interdisciplinary work in areas such as algebraic topology or analytic number theory can arise.  Professional scientists can be heard remarking about the ``completely different planets'' occupied by theoretical physics and mathematical physics-- distinctions that are often difficult to appreciate even to bright undergraduates, let alone a scientist from several centuries ago when even the most primitive concepts underlying modern physics were in their nascent stages of formalization.  \\

In the modern era, we have seen so many examples of mathematical success, both in the rich conceptual structures of pure mathematics, and in the application of mathematics to the physical sciences, that we can lose sight of the fact that there may have been a time when one might have been justified in maintaining some amount of doubt about the long term viability of mathematics itself as a worthy topic of investigation.  The world today has been so fundamentally imbued with the successes of mathematics, from the earth shattering precision of quantum electrodynamics and general relativity, to the society transforming potential of the Internet, medical imaging, and autonomous vehicles, that it is easy to forget to ask the question of why mathematics \emph{works at all} in describing natural phenomena.  And in an earlier era with far fewer examples of mathematical success in the natural sciences, what might have been the competing schools of thought that scientific minds might have otherwise invested their efforts in? \\

In the essay ``On the unreasonable effectiveness of mathematics in the natural sciences,'' physicist and quantum mechanics pioneer, Eugene Wigner, writes the following:
\begin{quote}
{\small Most of what will be said on these questions will not be new; it has probably occurred to most scientists in one form or another. My principal aim is to illuminate it from several sides. The first point is that the enormous usefulness of mathematics in the natural sciences is something bordering on the mysterious and that there is no rational explanation for it. Second, it is just this uncanny usefulness of mathematical concepts that raises the question of the uniqueness of our physical theories. In order to establish the first point, that mathematics plays an unreasonably important role in physics, it will be useful to say a few words on the question, ``What is mathematics?'', then, ``What is physics?'', then, how mathematics enters physical theories, and last, why the success of mathematics in its role in physics appears so baffling.\footnote{Eugene Wigner, ``On the unreasonable effectiveness of mathematics in the natural sciences,'' Communications in Pure and Applied Mathematics, vol. 13, No. I 1960.}}
\end{quote}

One interesting hypothesis that Wigner suggests is that the success of mathematics in contemporary science is partly due to a selective bias.  That is, as scientists, we have chosen, somewhat by necessity, to examine a subset of physical phenomena which are amenable to mathematical description.  In understanding how mathematical thought, and particularly its role in the physical sciences, could have taken root, it is worth examining the most general property of mathematics, namely as a symbolic alphabet for representing concepts.  What I will take as a starting point for the remainder of this discussion is the observation that the art of memory should also be thought of as a symbolic alphabet for representing concepts.  While this may not have been the perspective taken during antiquity, it is in the Renaissance transformation of the art of memory and the memory theater of Giulio Camillo that we see the worldview develop that mnemonic images could be used to represent the very structure of reality, in other words, \emph{as a potential basis for physical theories}.  Consequently, I suspect that in an era where the notion of mathematical modeling and symbolic representation were in their infancy, the distinction between the memory palaces of the mnemonic method and mathematics itself may have been viewed as quite small, or rather bearing the kind of resemblance that a modern mathematician might ascribe to analysis and algebra, or between a mathematical derivation in a physics paper and a computer simulation of the same system.  That is, these two forms of symbolic systems had enough in common that contrasting them and advocating the use of one over the other would have been a reasonable thing to do.  For things substantially more dissimilar, it would be completely unnecessary-- one doesn't need to advocate the use of a broom instead of an apple for sweeping the floor, but might yell at someone about to use a mop.  \\

I raise this point because I believe it provides us with a novel approach to attacking the question ``what is mathematics?'' or ``why can mathematics be used to describe physical phenomena?'' Suppose we would like to compare symbolic systems by the degree to which they possess two different qualities-- the capacity for aiding human memory and the capacity for building up successive layers of conceptual abstraction.  On such a scale, we would imagine that the symbols of the mnemonic method have an extremely high degree of memorability, but low capacity for abstraction, whereas the symbols of usual mathematics have a high degree of inherent potential for conceptual abstraction, but a moderately low inherent tendency toward precise recall.  It seems that human natural language lies somewhere in between the mnemonic method and mathematics on both scales (see Figure 2).  \\

\begin{figure}\label{fig:sym-sys}
\begin{center}
\includegraphics[width=.75\textwidth]{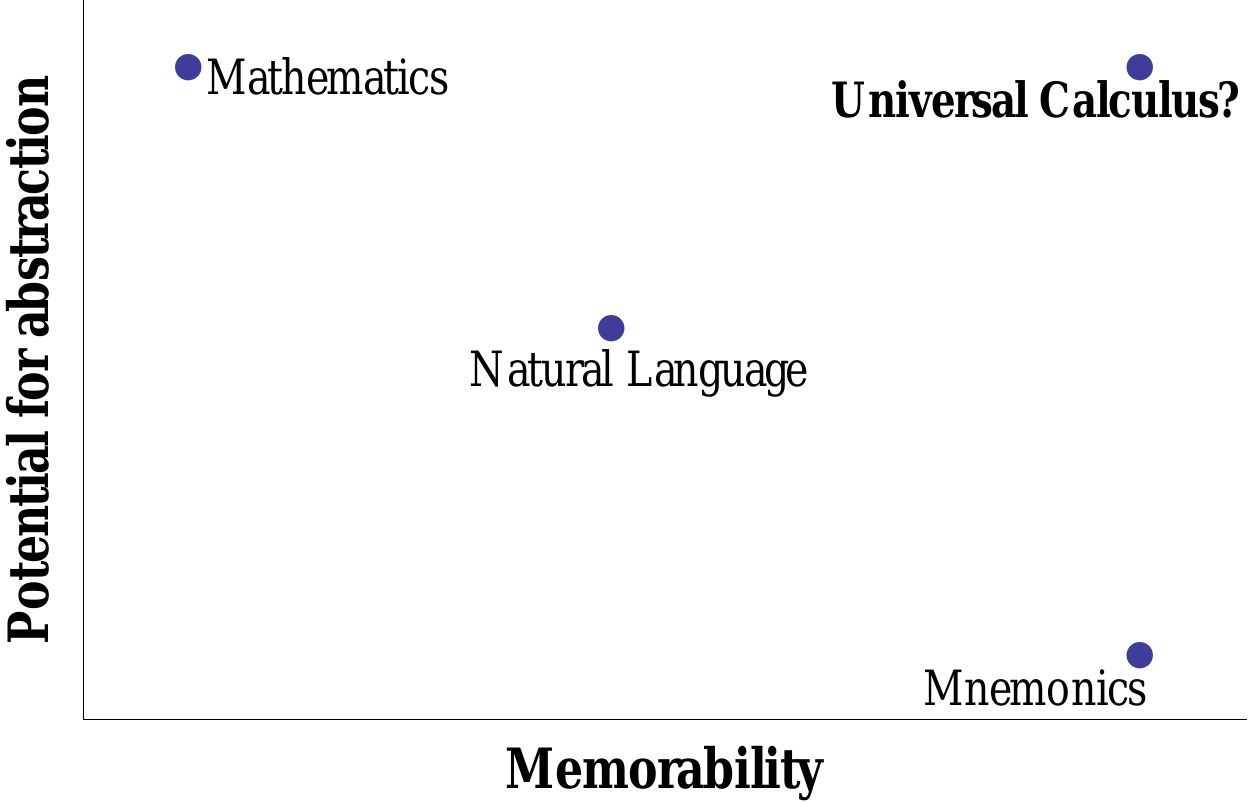}
\caption{{\small Cartoon diagram showing where different symbolic systems lie with respect to the criteria of memorability and potential for conceptual abstraction.  The point in the upper-right indicates a hypothetical system combining the high degree of memorability of mnemonics with the abstract potential of mathematics-- precisely what Leibniz was attempting to achieve with the universal calculus.}}
\end{center}
\end{figure}

We might then ask what other points on this graph could represent.  In particular, is it possible to develop a symbolic system which has the inherent capacity to aid the human memory, as the mnemonic method does, but which, like mathematics, also has the potential for building many successive layers of abstraction?  Indeed, I believe that this is exactly the question that Leibniz was tackling in formulating the notion of the universal calculus.  Recall that the agenda of the universal calculus was to first build an encyclopedia, in a manner akin to the dialectic method, and to then assign a symbol for each concept.  But Leibniz was keenly aware of the controversy surrounding the mnemonic method's problematic side effects for generating conceptual interference, and was looking to the Chinese and Egyptian alphabets as possible sources of inspiration for how he could retain the memory aiding features of the mnemonic method, while eliminating the unnecessary conceptual interference inherent in complicated and fantastic images. \\

I find Leibniz' vision to be truly romantic and beautiful, and yet, I also have an inherent skepticism that it would have worked.  Suppose, for example, that in the dialectic tradition, we create a structured list of all biological knowledge relevant to frogs.  As Ramus advises us, we organize the information starting with the most general and proceeding towards the most specific.  Now, following Leibniz, we attempt to create a symbolic representation for each of these concepts (recall the example given earlier of associating predicates with prime numbers).  Would we ultimately arrive at a ``calculus of frogs?''  This is difficult to imagine, and yet it seems that something along these lines is what Leibniz was trying to accomplish.  It is possible that enough has been forgotten of the thought process of the era that what I have written here is a misinterpretation of Leibniz' agenda.  In that case, it seems like ripe fodder for further historical work to re-examine the primary sources to understand why Leibinz, a man who had just created one of the most powerful mathematical structures the world has seen, had tremendous optimism that the universal calculus based on the hybrid mnemonic-dialectic method would have given rise to similar levels of conceptual abstraction and capacity for prediction as the infinitesimal calculus.  We see in the following inspired words, the zeal with which Leibniz pursued this vision:
\begin{quote}
{\small My invention contains all the functions of reason: it is a judge for controversies; an interpreter of notions; a scale for weighing probabilities; a compass which guides us through the ocean of experience; an inventory of things; a table of thoughts; a microscope for scrutinizing things close at hand; an innocent magic; a non-chimerical cabala; a writing which everyone can read in his own language; and finally a language which can be learnt in a few weeks, traveling swiftly across the world, carrying the true religion with it, wherever it goes.\footnote{Gottfried Wilhelm Leibniz, ed. Paul Ritter \emph{et al.}, \emph{S\"{a}mtliche Schriften und Briefe herausgegeben von der Preussischen Akademie der Wissenschaften} (Darmstadt, 1923-), I, ii, pp. 167-169, quoted in Rossi, p. 191.}}
\end{quote}

Assuming for the moment that Leibniz' hybrid mnemonic-dialectic method is unlikely to have given rise to a universal calculus in the way his more mathematical work gave rise to the integral and differential calculus, let me now state a question that I believe to be an alternative approach to attacking the questions ``what is mathematics?'' or ``why can mathematics be used to describe natural phenomena?''  The question is the following:  why does mathematics keep going where Leibniz' hybrid mnemonic-dialectic method stops?  That is, it appears as though both mathematics and mnemonics share the core property of being symbolic representations of concepts, however, the latent potential in mathematics for building successive layers of abstraction is infinite, whereas with mnemonics, even when used in conjunction with the logical structure of a dialectically organized encyclopedia, it seems as though it doesn't quite get off the ground.  And yet for some of the greatest minds of the scientific revolution-- Gottfried Leibniz in particular, but also Rene Descartes-- there was a tremendous amount of confidence that mnemonics would give rise to precisely the kind of conceptual power that mathematics has ultimately given us, and for which there was only a small amount of evidence at the time.  \\

There is one obvious hypothesis for why mnemonics lack the abstract potential of mathematics, which I think ultimately fails, but is worth examining anyway.  As I have stated throughout this essay, one problematic feature of mnemonics is the built-in conceptual interference that the techniques give rise to.  In fact, it seems as though conceptual interference is precisely what enables the powerful ability to recall minutiae with such accuracy.  Suppose I want to remember that the adjective ``feline" means ``possessing the qualities of a cat."  I can create a mnemonic that uses the sound of the prefix ``fe" to remind me of the element iron, and I can imagine a huge iron cat, possibly two huge iron cats standing on either side of a large iron gate which stands at the entrance to a large palace, and in which many cats are standing in ``line'' to visit.  Using this association, I can remember that ``feline" means ``possessing the qualities of a cat."  However, for the most part, the mnemonic itself consists of information largely irrelevant to anything related to cats, and so in a fundamental way, this image is incredibly misleading, although our brains seem to possess the remarkable ability to know to ignore the vast majority of the information contained in the mnemonic image.  I am unlikely to begin to think that cats are able to stand in line or that there are palaces devoted to cats by virtue of using this particular mnemonic. \\

So it seems as though mnemonics gain their power by taking advantage of obscure and fantastic connections between word and object and that while this may aid the memory, it may also be what inhibits further conceptual potential.  In fact, as I have described earlier in the article, this objection is precisely what was raised by Ramus and his disciples, the proponents of the dialectic method.  Of course, their objection was not entirely an intellectual one-- it was that the use of the mnemonic method would pollute the mind with unclean images and that this would corrupt one's spiritual character.  But it seems that for other intellectuals, there would have been completely secular reasons for opposing use of the mnemonic method, and conceptual interference was the basis for this objection.  \\

Can we conclude then that the reason that mnemonics do not reach the conceptual capacity of mathematics is that the information content of mnemonics is dominated by concepts peripheral to the core notion intended to be remembered?  This is possible, however, it seems worth understanding why such a phase transition would happen, because mathematics itself does contain some amount of conceptual interference, albeit substantially less.  For example, in quantum mechanics, we frequently denote the system density matrix with the symbol $\rho$.  The symbol is round-- does that mean the density matrix itself is also round?  Of course, not, it is merely a symbol and could just as well be represented by $\psi$, $\Gamma$, or $\otimes$.  Probability distributions are often times denoted by $\pi$.  Does that mean that they have some deep connection with the ratio of a circle's circumference to its diameter?  Again, of course, not, but it is clear from this example that mathematical notation can also give rise to conceptual interference, albeit substantially less than the elaborate scenery of the mnemonic method.  \\

Before bringing this discussion to a close, let me restate my primary purpose in this section.  For modern intellectuals, Leibniz' agenda of constructing the universal calculus via the mnemonic-dialectic method sounds extremely strange.  How could mnemonic images possibly give rise to either conceptual abstraction or predictive power when they are simply one-to-one maps from some list of facts?  And yet it appears as though this viewpoint had quite a bit of support during the 17th century.  In particular, it is noteworthy that Leibniz, who stands among elite company in having developed the infinitesimal calculus, had the most confidence of any thinker of the era-- it would be surprising if the obvious objections to this project were lost on him.  Therefore, I believe that the question ``why does mathematics have a near infinite capacity for conceptual abstraction whereas mnemonics do not?'' is a new one and may provide a fresh perspective on how to attack some long-standing questions in the foundations of mathematics.  Perhaps in trying to articulate explicitly the many implicit assumptions in our worldview, we will discover something novel and of practical relevance to contemporary intellectual questions.  

\subsection*{Acknowledgements}
I would like to thank Michael Rosen, Aaswath Raman, and Doug Bemis for insightful discussions and critical reading of the manuscript.  

\end{document}